Title: Beam Dynamics for ARIA

Author(s): Ekdahl, Carl August Jr.

Intended for: Report

Issued: 2014-10-14 (rev.1)

# Beam Dynamics for ARIA

Carl Ekdahl

*Los Alamos National Laboratory*

*Abstract*

Beam dynamics issues are assessed for a new linear induction electron accelerator being designed for flash radiography of large explosively driven hydrodynamic experiments. Special attention is paid to equilibrium beam transport, possible emittance growth, and beam stability. It is concluded that a radiographic quality beam will be produced possible if engineering standards and construction details are equivalent to those on the present radiography accelerators at Los Alamos.

**I. INTRODUCTION**

The Advanced Radiography Induction Accelerator (ARIA) is an electron linear induction accelerator (LIA) for flash radiography of explosively driven hydrodynamic experiments. Flash radiography of such experiments is a time proven diagnostic in use world-wide [1] [2], and ARIA is the most recent accelerator designed for this purpose. ARIA is designed to meet the initial flash radiography requirements:

- Two pulses on a common axis to enable velocity measurements.
- Pulse spacing variable from 200 ns to 3000 ns to accommodate different experiments.
- Each bremsstrahlung radiation pulse less than 50-ns full width at half maximum (FWHM) to minimize motion blur.
- Dose per pulse on axis at 1 m variable from 40 to 150 rad(Pt) for sufficient signal to noise ratio on the radiograph.
- End-point energy of 12-MeV to ensure that there is enough useful dose in the energy range of maximum penetrability of the object.
- Spot size less than 0.7-mm FWHM for adequate resolution of details.

The ARIA accelerator is similar to the single-pulse LIA at the Dual Axis Radiography for Hydrodynamic Testing (DARHT) facility at Los Alamos National Laboratory.

Our first iteration of the ARIA accelerator design is code named "Wagner." The Wagner design has a 2-kA, 3-MeV injector and 36 cells, each producing a 250-keV beam-loaded accelerating potential, for a final energy of 12 MeV [3]. The injector uses a hot dispenser cathode to ensure multiple-pulse operation without the gap closure typical of high-current cold cathodes.

Cell design is based on DARHT Axis-I, but with the ferrite cores replaced with enough Metglas to eventually provide four pulse operation. The cavity shape and materials of the Wagner cell are identical to the Axis-I cell in order to have the same BBU properties. Except for cell dimensions, the physical layout of the accelerator is the same as Axis-I, with cells grouped in blocks of four, and pumping stations between blocks of eight. The general layout of ARIA is shown in Fig. 1.

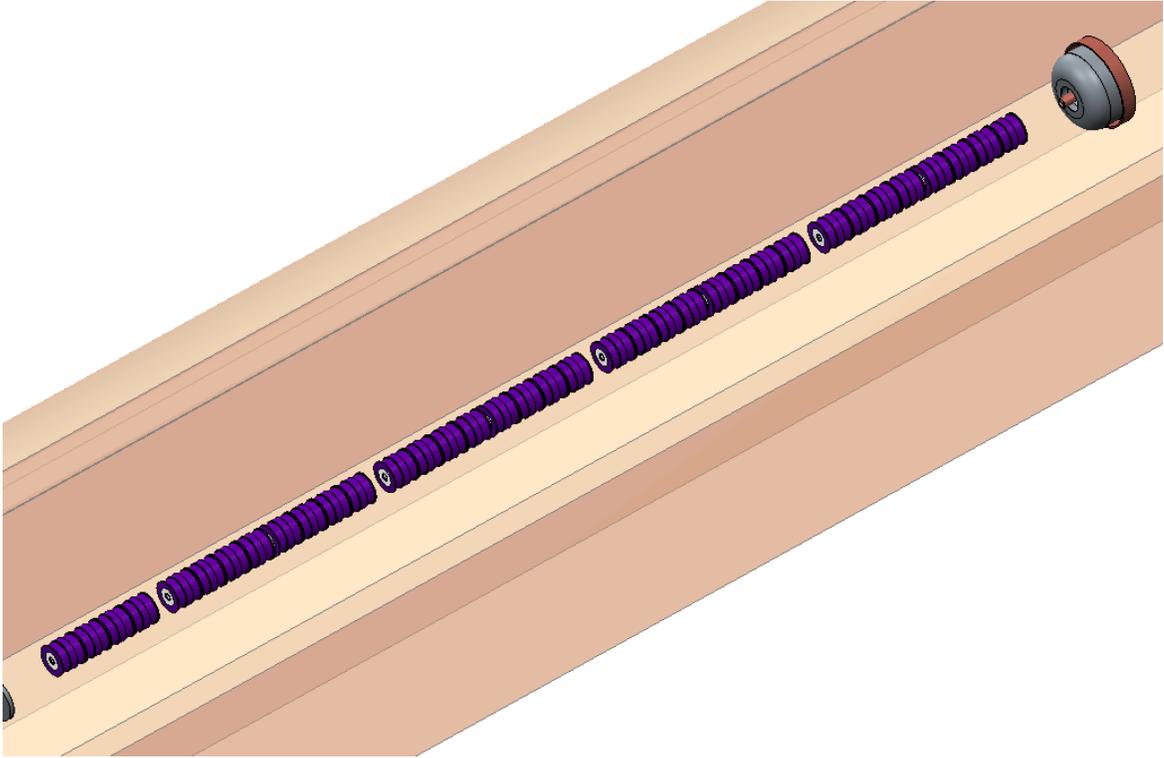

Figure 1: Major component layout of the ARIA advanced radiography accelerator, showing the injector diode and the accelerator induction cells.

The electron beam from DARHT Axis-I is of sufficient quality that the resulting bremsstrahlung radiation source spots exceeds all anticipated requirements for hydrodynamic testing. However, there are enough differences between ARIA and Axis-I that an assessment of beam dynamic issues in ARIA is called for. These issues include beam transport, motion, stability, and emittance. Effective management of these issues has consequences for accelerator engineering choices. An initial investigation of these issues based on a preliminary design for ARIA, and how these influence accelerator engineering, is the purpose of this article.

We are using all of the tools at our disposal to assess the beam dynamics issues on ARIA:

- Experimental data from DARHT Axis-I and Axis-II
- Physics simulation codes
- Analytic theory

This article relies heavily on results of simulation codes such as a suite of 2D and 3D of electromagnetic and charged particle tracking codes, beam envelope codes, and particle-in-cell (PIC) codes.

The article is organized as follows. Generation of a realistic 2-kA, 3-MeV beam to use for initial conditions for accelerator computations is discussed in Section II. Transport of this beam through the accelerator, including potential for emittance growth is covered in Section III. Beam stability issues are explored in Section IV. Section V is an initial account of the impact of beam dynamics on accelerator engineering.

**II. INJECTOR**

Simulations of beam dynamics in the ARIA LIA require the beam parameters at the point of injection. Since the ARIA injector has not yet been designed, some assumptions about these parameters must be made. In order for the assumed parameters to be as realistic as possible, I created a model of a hypothetical injector, and simulated the beam it produced with a finite-element e-gun design code. The parameters of the beam produced by this injector were then used as initial conditions in all subsequent simulations.

I made several assumptions about the final injector design, and then modeled it using the Field Precision Tricomp suite of finite-element electron gun design codes [4] [5]. The basic requirements for the beam injected into the LIA are multiple 50-ns, 3-MeV, 2-kA pulses separated by more than 100 ns.

A heated dispenser cathode was assumed to be the source of electrons to avoid possible impedance collapse issues associated with cold cathodes. I reduced the size of the cathode from that of the DARHT Axis-II cathode in order to reduce radiative heating of diode components, and to improve beam optics (compared to Axis-II). The 6.5-inch Axis-II cathode provides ~1.7 kA at ~2.2 MV diode voltage when heated to ~1100 C brightness temperature. This is only ~8 A/cm$^2$, although 311M cathodes are capable of better than twice this at the same temperature (612M and 411M are rated at even higher emissivity) [6]. For the ARIA diode design I assumed a cathode with a 5-inch diameter, which would need to emit 15.8 A/cm$^2$ for the 2-kA beam.

I assumed an external bucking coil like on Axis-I, instead of one interior to the cathode shroud as on Axis-II. This minimizes the loads on the cathode stalk, eliminates the need for extensive cooling and heat shielding, and provides minimum sensitivity of flux null to cathode position errors or drive current deviations [7]. For these simulations, I also used the Axis-I anode solenoid and the same locations of these two magnets with respect to the cathode as on Axis-I.

The outer wall of the diode vacuum region was set at the same as Axis-I to accommodate a conservative radial insulator.

For reference, simulation of the Axis-I diode with a 2-inch diameter emission surface collocated with the flat cathode shroud, and a 7-inch cathode-to-anode (shroud) gap[*] produces ~2 kA with 4 MV applied. With 4 MV the maximum field on the cathode shroud is ~300 kV/cm (on the curved surface). Therefore, the targets for ARIA are 2-kA at 3 MV with less than 300 kV/cm on the cathode shroud.

---

[*] The cathode emission surface to anode shroud dimension should not be confused with the actual anode-cathode (AK) gap, which is defined by a virtual anode located in the exit pipe somewhat downstream of the anode shroud. However, it is a convenient dimension for diode simulation comparisons, so it is frequently called the "AK gap," and that common usage is preserved in this article.

This beam is fully relativistic, so the AK gap spacing for ARIA was initially estimated using modified Jory-Trivelpiece (JT) scaling [8]. For an infinite planar diode, JT scaling predicts the product $Jd^2$ as a function of diode voltage. (Here $J$ is the current density at cathode, and $d$ is the AK gap.) However, the Axis-I and ARIA diode geometries are far from infinite planar, so the JT scaling was only used as a starting point for empirically determining the proper gap spacing for the diode.

Finally, I added Pierce focusing electrodes to focus the large 5-inch diameter beam at the emitter down into the 6-inch diameter exit pipe. This will allow the use of the same ~6-inch diameter beam pipe at the anode as Axis-I, even with the substantially larger cathode surface (5-inch diameter compared with Axis-I 2-inch diameter). Pierce electrodes also prevent the hollow beam that is characteristic of flat cathodes [9]. Under certain conditions, hollow beams are unstable to the diocotron instability when focused by strong solenoidal magnetic fields [10] [11]. The diocotron instability produces azimuthal inhomogeneity, which increases emittance, so I assumed that we would want to prevent that in ARIA.

The classic Pierce electrode geometry with the focusing electrode intersecting the emitting surface has a strong spherical aberration near the edge, which causes some edge focusing of the beam. In order to mitigate this I added a small (0.2 inch) annulus of flat, non-emitting surface around the cathode.

Simulations of this diode produced 2.06 kA at 3 MV with an 8.2-inch gap. The maximum field on the cathode shroud was 224 kV/cm, which is substantially less than presently on the Axis-I cathode shroud. The uniform beam was focused to ~2.7 cm rms radius at the anode pipe entrance (Fig. 2 and Fig. 3). Initial conditions for envelope and PIC simulations were extracted at a position 54-cm from the cathode, where the beam size is at a local maximum (the best place for launching the PIC code [12]). Moreover, this position is far enough into the beam pipe that axial accelerating fields from the diode are negligible (, a necessary constraint on the initial position used in our envelope equation code [13] At this location, with 120 A on the anode magnet, the beam equivalent envelope radius was 4.2 cm, and it had a 693 π-mm-mr normalized emittance. This emittance seems high compared with that calculated for the large cathode Axis-II diode, so it should be checked with PIC (LSP) simulations. Therefore, in subsequent LIA beam dynamics simulations I used these TRAK simulation results for beam size and zero convergence, but varied the emittance from 200 π-mm-mr to 800 π-mm-mr to cover a range of possibilities.

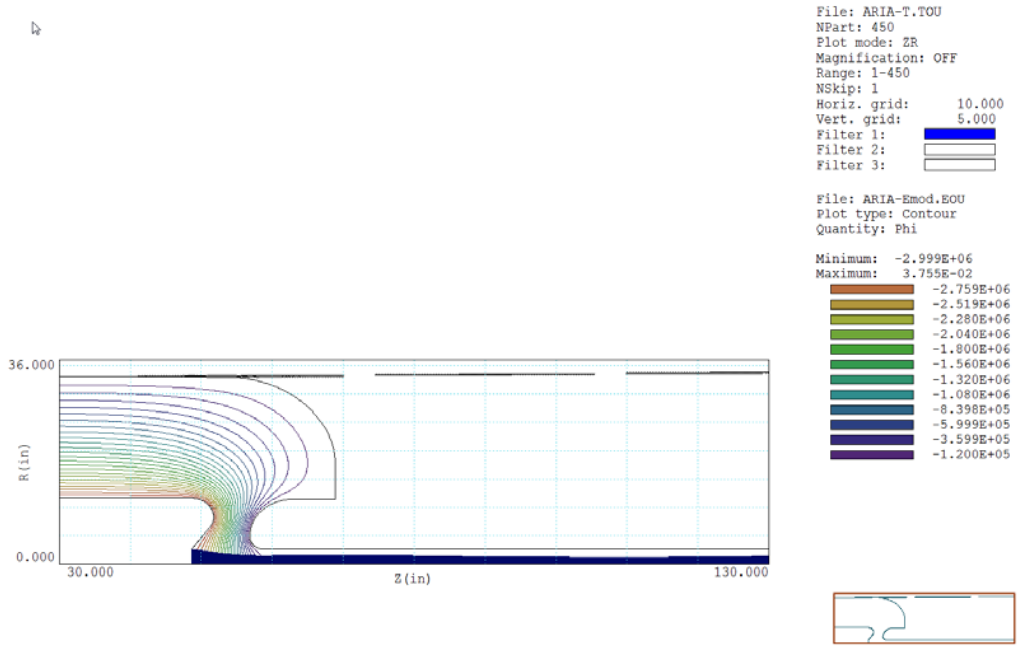

Figure 2: TriComp simulation of the 2.05- kA space-charge limited beam produced by a 3-MV diode injector that has a 5-inch diameter dispenser cathode electron source. (Dimensions are in inches)

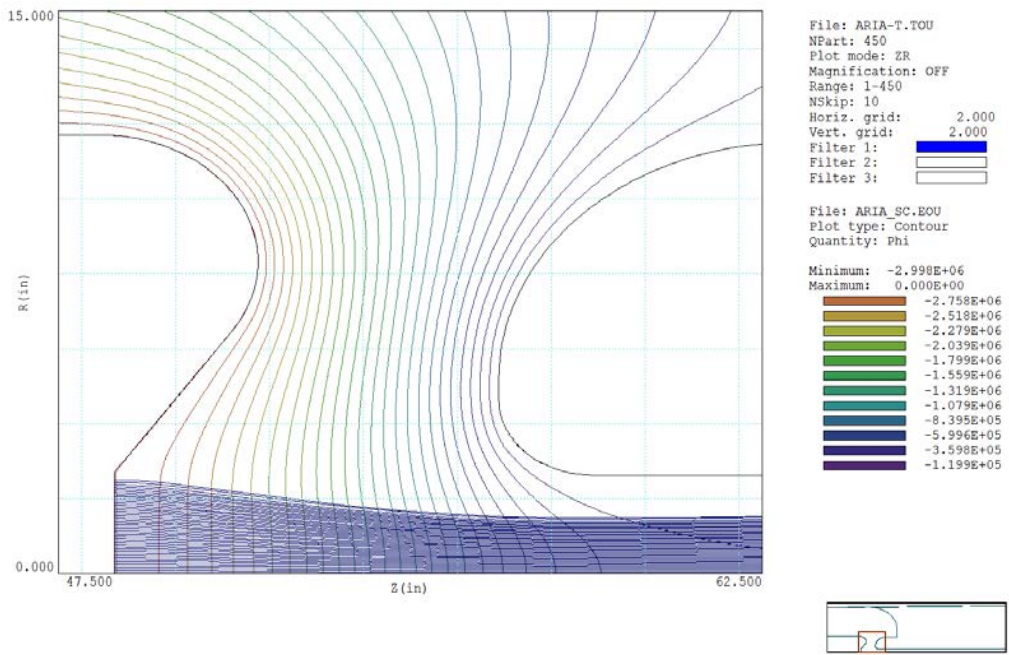

Figure 3: Detail of A-K region of TriComp simulation of the 2.06- kA space-charge limited beam produced by a 3-MV diode injector that has a 5-inch diameter dispenser cathode electron source. Equipotential contours include the beam space charge. (Dimensions are in inches)

## III. BEAM TRANSPORT

The electron beam is transported through the LIA using solenoidal magnetic focusing fields. This is an efficient and convenient means that has been used in all electron LIAs since the very first. Each accelerating cell has a solenoid incorporated into it, as well as dipole windings to produce orthogonal transverse steering fields. The magnetic field produced by these magnets is called the tune of the accelerator. This section reports the results of beam simulations of tunes for beam transport through the ARIA LIA.

### A. *External electromagnetic fields*

*Magnetic fields*

The solenoidal magnetic fields used for the beam simulations are based on the ARIA Wagner cell design [14], shown in Fig. 4. The solenoid is a double layer of square, hollow core conductor with 156 total turns. I used the Field Precision PerMag program [5] to calculate the field in the cell for a 100 A drive current. Figure 5 shows the resulting magnetic field. The field on axis was fit with the model used in our XTR envelope code [15]. The simulation indicates that this magnet will produce a peak field on axis of 3.64 Gauss/Amp. Figure 6 is a plot of the axial magnetic flux density, $B_z$ , on axis as calculated by PerMag, along with the XTR model fit for comparison.

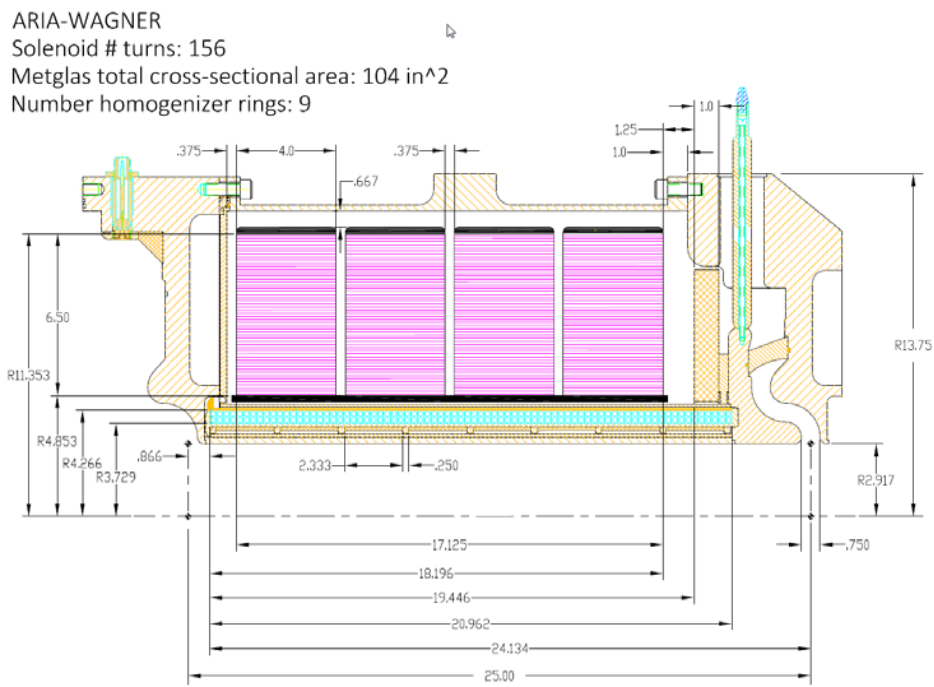

Figure 4: ARIA cell design for Wagner.

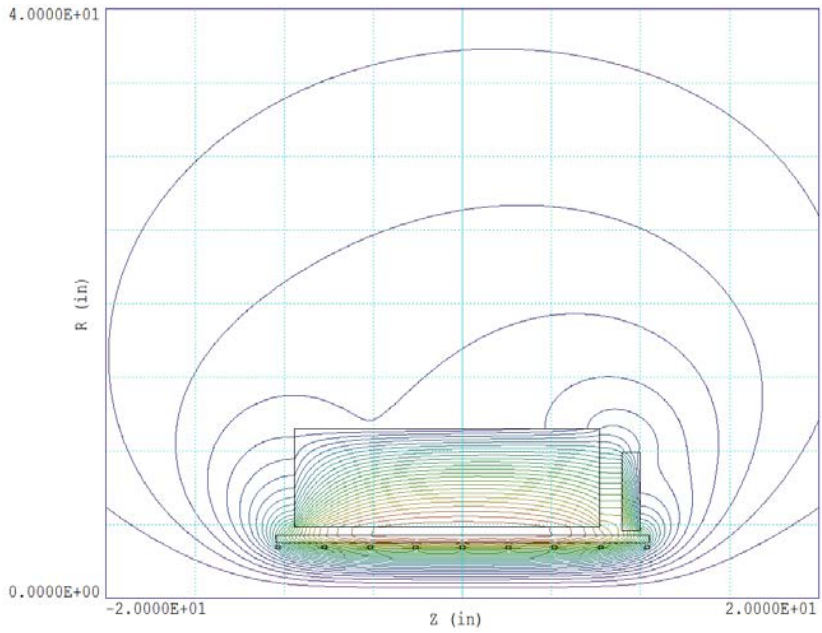

Figure 5: Contours of $rP_\theta$ for the ARIA-Wagner cell.

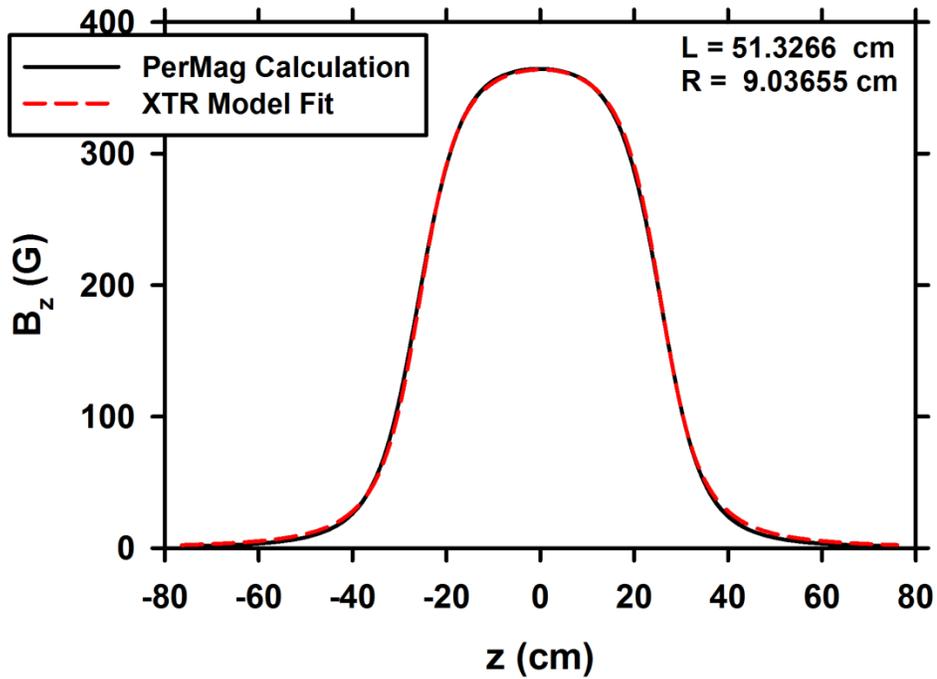

Figure 6: Axial magnetic field on the axis of the ARIA cell solenoid powered with 100 A.. (solid black line) Calculated with the TriComp PerMag code. (dashed red line) Model fit for use in XTR envelope code. (The model is an ideal sheet solenoid with 51.3266 cm length and 9.03655 radius.)

*Electric fields*

The XTR envelope code uses a thin Einzel-lens approximation for the acceleration and focusing of the beam by the LIA gaps, but the PIC code requires explicit $E_z$ on axis in tabular form. In order to provide this table for ARIA, I performed an electrostatic simulation of the gap region for the Axis-I geometry, which will be duplicated on ARIA. The TriComp ESTA code was used for this calculation. Figure 7 shows the electrostatic potentials for this simulation with 250 kV across the gap. Only the features of the gap region that might affect the field on axis were included. Figure 8 shows the resulting field on axis, which was used with the gap locations to create an input file for LSP-Slice PIC simulations. The functional form of $E_z(z)$ from ESTAT (Fig. 8) compares well with the table used in previous PIC simulations of Axis-II, once the difference in beam pipe radius is accounted for.

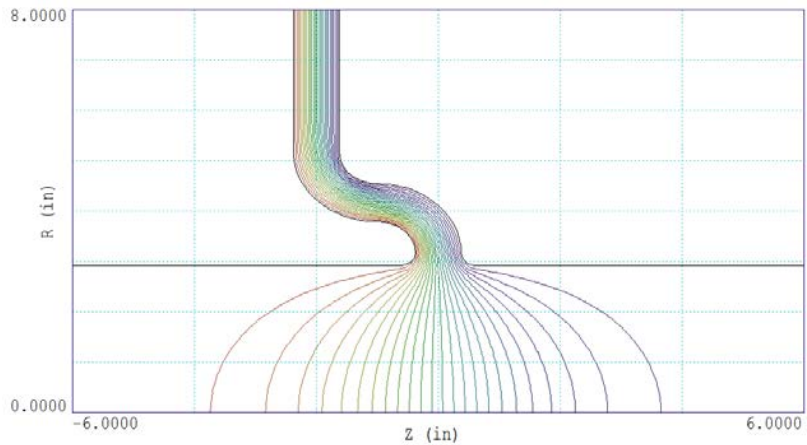

Figure 7: Equipotentials of the accelerating electric field at 10-kV intervals in the region of the ARIA gap for 250-kV gap voltage. Simulation was performed using the TriComp ESTAT code.

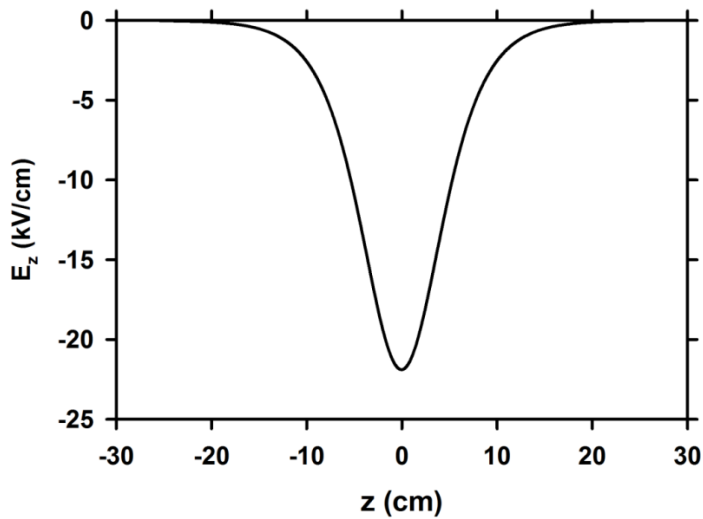

Figure 8: Accelerating electric field on axis calculated by ESTAT for the ARIA cell.

## B. *Matched Beam Tune*

Design of tunes for the DARHT accelerators is accomplished using envelope codes. The two most frequently used are XTR and LAMDA. XTR was written by Paul Allison in the IDL language [6]. LAMDA was originally written by Tom Hughes and R. Clark [7]. In both of these codes the radius $a$ of a uniform density beam is calculated from an envelope equation [8]. In the DARHT accelerators the beam is born at the cathode with no kinetic angular momentum and a reverse polarity solenoid to cancel out the magnetic flux. Thus, the beam has no canonical angular momentum, and the envelope equation is

$$\frac{d^2r}{dz^2} = -\frac{1}{\beta^2\gamma}\frac{d\gamma}{dz}\frac{dr}{dz} - \frac{1}{2\beta^2\gamma}\frac{d^2\gamma}{dz^2}r - k_\beta^2 r + \frac{K}{r} + \frac{\varepsilon^2}{r^3} \tag{1}$$

It can be shown that this same equation holds true for any axisymmetric distribution [9], so long as the radius of the equivalent uniform beam is related to the rms radius of the actual distribution by $r = \sqrt{2}R_{rms}$. Here, $\beta = v_e/c$, $\gamma = \sqrt{1-1/\beta^2}$, are the usual relativistic parameters, and the beam electron kinetic energy is $KE = (\gamma-1)m_e c^2$. The betatron wavelength is

$$k_\beta = \frac{2\pi B_z}{\mu_0 I_A} \tag{2}$$

where $I_A = 17.08\beta\gamma$ kA, and the generalized perveance is $K = 2I_b/\beta^2\gamma^2 I_A$. The emittance which appears in Eq. (1) is related to the normalized emittance by $\varepsilon = \varepsilon_n/\beta\gamma$, where

$$\varepsilon_n = 2\beta\gamma\sqrt{\langle r^2\rangle\left[\langle r'^2\rangle + \langle (v_\theta/\beta c)^2\rangle\right] - \langle rr'\rangle^2 - \langle rv_\theta/\beta c\rangle^2} \tag{3}$$

which is invariant through the accelerator under certain conditions.

The simple envelope equation in Eq. (1) is further improved in XTR as follows [10]. The energy dependence of the beam due to the gaps is approximated by a linear increase in $\gamma$ accompanied by a thin-einzel-lens focus. Between gaps $\gamma$ used in Eq. (1) is the value at the beam edge, which is space-charge depressed by $\Delta\Phi \approx 30 I_b (2\ln R_w/r)$, where $R_w$ is the radius of the beam pipe [8]. XTR also uses the magnetic field at the beam edge, including a first order approximation to account for the flux excluded by a beam rigidly rotating in the magnetic field due to the invariance of canonical angular momentum [11].

The XTR envelope code was used to develop tunes for the ARIA Wagner accelerator design shown in Fig. 1. Because of the success of DARHT Axis-I, I initially developed a tune similar to the nominal Axis-I tune that has been frequently used with a 2-inch cathode (Fig. 9). A low-field tune for ARIA that is comparable to the nominal Axis-I tune is shown in Fig. 10. This tune uses only modest magnetic fields, which is beneficial from the power consumption and heat management standpoint. On the other hand, higher magnetic fields may be required to defeat the BBU instability, to be discussed in a following section. For this purpose, I also developed a high-field tune.

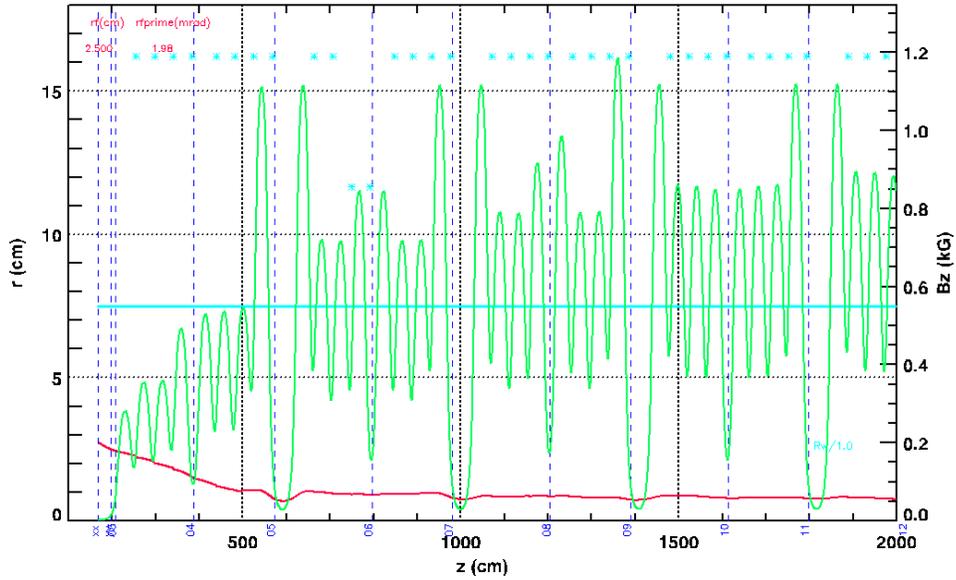

Figure 9: Nominal tune for DARHT Axis-I. (red) Beam envelope. (green) Magnetic guide field on axis, right scale. (solid cyan) Beam pipe wall. (cyan asterisks) Accelerating cell potential. (blue dashed) BPM locations.

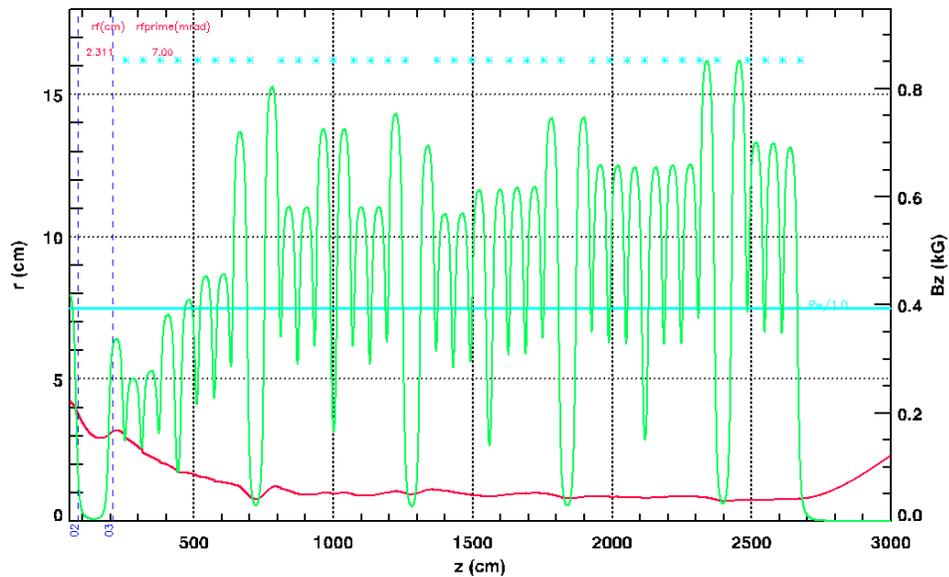

Figure 10: Low guide-field tune for ARIA Wagner. (red) Beam envelope. (green) Magnetic guide field on axis, right scale. (solid cyan) Beam pipe wall. (cyan asterisks) Accelerating cell potential.

The floor plan for the ARIA Wagner accelerator has large gaps between eight-cell blocks to accommodate pumping and alignment bellows. I added a large inter-cellblock coil in this region to bridge the gap, so that it could be tuned to a high enough magnetic field to effectively suppress the BBU. The inter-cellblock magnet model that I used has 100 turns (4 layers) with a 10-cm effective length at an effective radius of 34 cm, having a peak field ~1.9 G/A (approximately the same conductor density as the cell solenoid). It is centered 15 cm downstream of the last accelerating gap in the cellblock. The high-field tune using three inter-cellblock solenoids is shown in Fig. 11. Obviously, it could be improved by further reducing the envelope oscillations, but it is adequate for demonstrating that BBU can be greatly suppressed on ARIA by using high magnetic fields. On the other hand, a high field tune with equivalent BBU growth, but lacking inter-cellblock magnets is desirable.

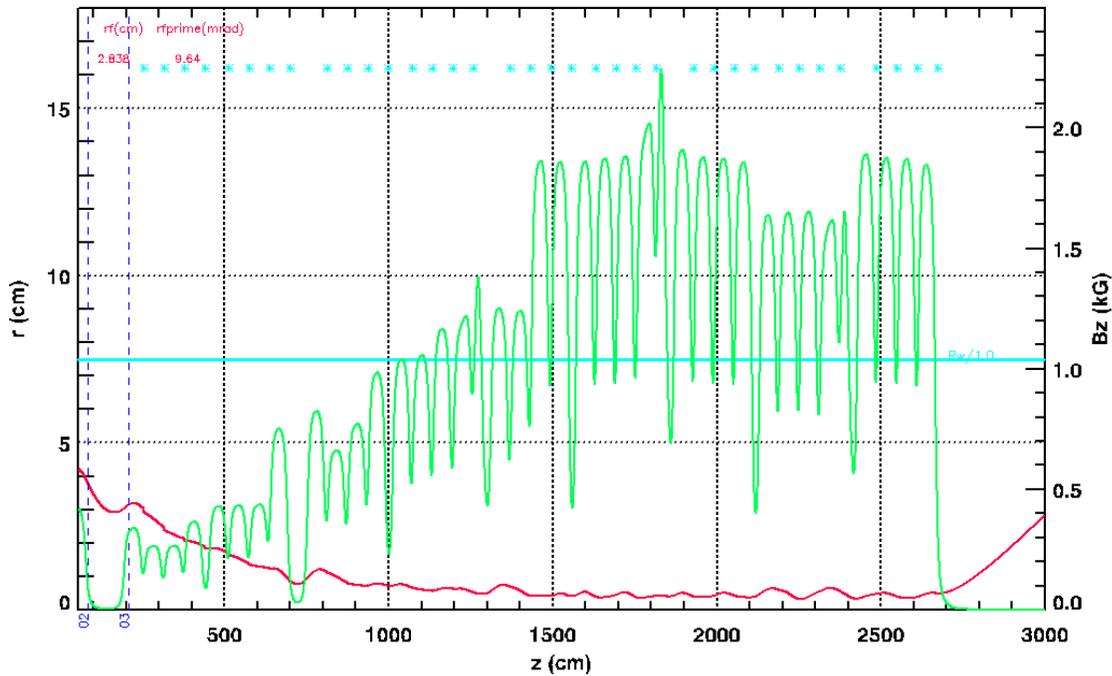

Figure 11: High guide-field tune for ARIA Wagner. (red) Beam envelope. (green) Magnetic guide field on axis, right scale. (solid cyan) Beam pipe wall. (cyan asterisks) Accelerating cell potential. Note the presence of inter-cellblock solenoids in gaps beyond z = 1000 cm.

## C. *Beam Emittance*

Emittance growth can result from envelope oscillations caused by a mismatch of the beam to the magnetic transport system. A badly mismatched beam exhibits large envelope oscillations, sometimes called a "sausage," "m=0," or "breathing" mode. The free energy in these oscillations feeds the growth of emittance [16]. The detailed mechanism of this contribution to emittance growth is parametric amplification of electron orbits that resonate with the envelope oscillation, expelling those electrons from the beam core into a halo [17] [18]

Beam emittance growth in the ARIA LIA was assessed using a particle-in-cell (PIC) computer code. It is based on the Large Scale Plasma (LSP) PIC code [19]. The LSP-slice algorithm is a simplified PIC model for steady-state beam transport in which the paraxial approximation is assumed [20]. A slice of beam particles located at an incident plane of constant z are initialized on a 2D transverse Cartesian ($x,y$) grid. The use of a Cartesian grid admits non-axisymmetric solutions, including beams that are off axis. Axisymmetric beam simulations were performed using a faster version of the code based on a 1D cylindrical grid. (Excellent agreement between the 2D and 1D results were obtained in comparison tests.)

The initial particle distribution of the slice is extracted from a full $x, y, z$ LSP simulation. The distribution is a uniform rigid rotor with additional random transverse velocity. The rotation is consistent with zero canonical angular momentum in the given solenoidal magnetic field at the launch position. The random transverse velocity is consistent with the specified emittance.

External fields are input as functions of $z$, and are applied at the instantaneous axial center-of-mass location. External fields that are azimuthally symmetric (fields from solenoids and gaps) are input as on-axis values, and the off-axis components are calculated up to sixth order using a power series expansion based on the Maxwell equations [21]. In this way the nonlinearities of the accelerator optics are included in the slice simulations. The on axis magnetic field ($B_z$) for the PIC simulations was from the XTR simulations. The on axis electric accelerating field ($E_z$) was obtained using the ESTAT gap model at each gap location.

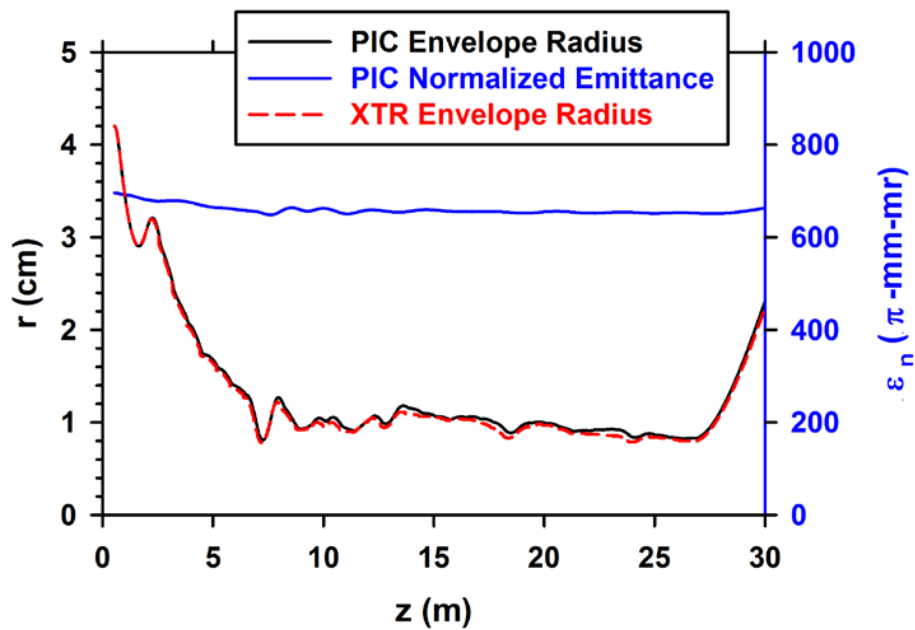

Figure 12: PIC simulation of ARIA low magnetic guide field tune. Black curve: ARIA beam envelope calculated by PIC code. Blue curve: Beam emittance calculated by PIC code. Also shown for comparison is the envelope calculated by XTR (Red dashed curve).

Using the XTR initial conditions for the PIC simulations the Wagnerian ARIA low-field tune produced no emittance growth (Fig. 12). As seen in Fig. 12, the envelope radius calculated by the PIC simulation compared very closely to the radius calculated the XTR envelope code. Large variations in the initial conditions produced mismatched, oscillating envelopes. Nevertheless, the emittance growth was less than ~18% over the entire range of initial conditions tried, so it appears that this tune is robust to emittance growth resulting from envelope oscillations.

Of greater concern is emittance growth in the high-field tune developed to suppress BBU. The envelope simulation of this tune exhibits a number of envelope oscillations, which is a concern for emittance growth. However, the emittance did not grow in the PIC simulation (Fig. 13). Again, the envelope radius from the PIC simulation compared very closely to the radius from the XTR envelope code, which is reassuring corroboration of the codes.

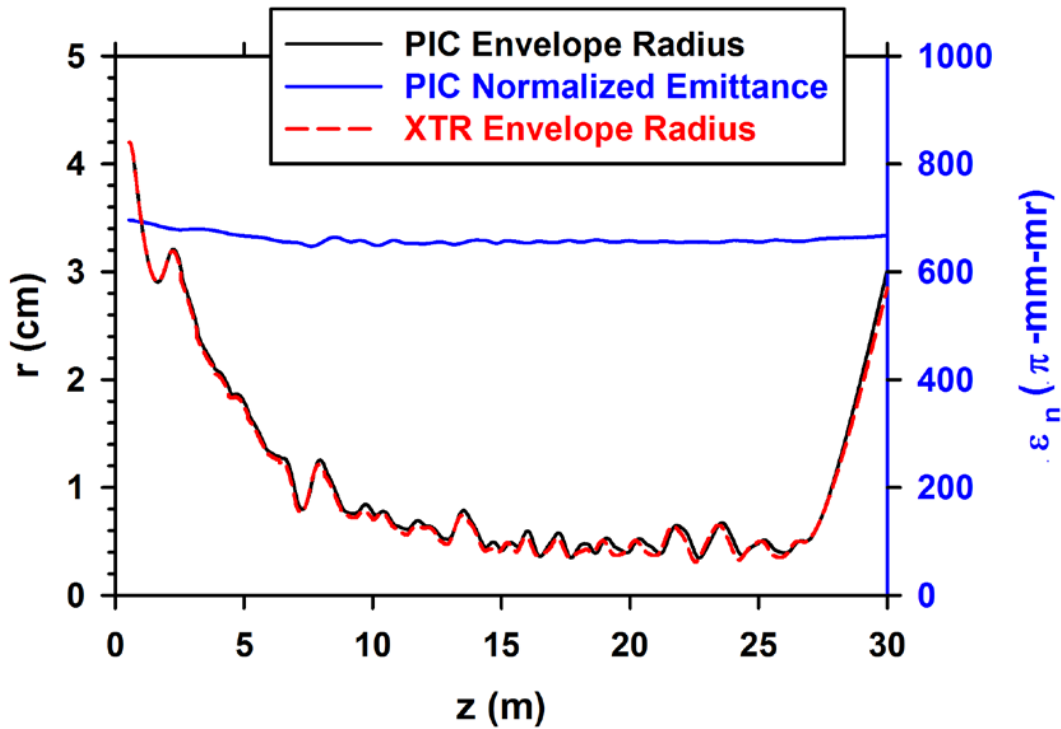

Figure 13: PIC simulation of ARIA high magnetic guide field tune. Black curve: ARIA beam envelope calculated by PIC code. Blue curve: Beam emittance calculated by PIC code. Also shown for comparison is the envelope calculated by XTR (Red dashed curve).

**D.** *Magnetic Field Crosstalk*

An effect in LIA accelerating cells that is seldom, if ever, mentioned is the crosstalk between the axial and azimuthal magnetic fields in the Metglas [22]. This is a result of the nonlinearity of the magnetization. Saturating the azimuthal field with the core reset pulse can effectively partially demagnetize the axial field in the material. The effect is equivalent to an apparent change in permeability for the axial field component. However, this has little influence on the focusing field on axis, which is dominated by the air path of flux outside of the core. Indeed, the focusing field on axis is almost entirely unaffected by large changes in the permeability of core material. For example, in the PerMag simulations of the ARIA Wagner solenoid a 10% change in Metglas permeability resulted in less than 0.01% change in field on axis.

## IV. BEAM STABILITY

In this section beam stability in the ARIA LIA is assessed. The two most worrisome problems are the beam breakup (BBU) instability and corkscrew motion, and they are discussed first. The diocotron instability is discussed next. If the injector produces a hollow beam, the diocotron instability may significantly increase emittance, especially if the injected beam has low energy. Finally, three other instabilities are briefly examined, but are not likely to cause problems: the parametric envelope instability, the resistive wall instability, and the ion hose instability.

### A. *Beam Breakup*

The most dangerous instability for electron linacs is the beam breakup (BBU) instability [23] [24]. For radiography LIAs it is particularly troublesome, because even if it is not strong enough to destroy the beam, the high-frequency BBU motion can blur the source spot , which is time-integrated over the over the pulselength. In a fast risetime LIA such as Axis-I or ARIA, BBU excited by the sharp beam head grows to a peak and then decays [24] (unlike on the slowly rising beam of Axis-II, where BBU grows from noise and corkscrew throughout the pulse [25]). The maximum amplitude of the BBU in high-current LIAs has been experimentally confirmed to be the theoretically predicted value [25] :

$$\xi(z) = \xi_0 \left[ \gamma_0 / \gamma(z) \right]^{1/2} \exp(\Gamma_m) \tag{4}$$

where subscript zero denotes initial conditions, and $\gamma$ is the relativistic mass factor. The exponent in this equation is [24]:

$$\Gamma_m(z) = \frac{I_{kA} N_g Z_{\perp \Omega/m}}{3 \times 10^4} \left\langle \frac{1}{B_{kG}} \right\rangle \tag{5}$$

where $\langle \ \rangle$ indicates an average over $z$, and $N_g$ is the number of accelerating gaps. Following experimental validation, this formulation was implemented in our XTR envelope code, and used to design Axis-II tunes that suppress BBU amplification to acceptable levels [26] [27].

The peak BBU is reached in the time

$$\tau_p = 2 \frac{Q \Gamma_m}{\omega_0} \tag{6}$$

after the arrival of the beam head [24], where $Q$ is the cavity quality factor and $\omega_0$ is the mode frequency.

For ARIA, we will use the exact gap and cavity geometry as Axis-I to ensure that BBU can be suppressed to the same amplitudes by making the transverse impedance the same. Using a displaced rod to excite it, the Axis-I transverse impedance was measured on a single cell to be 670 $\Omega/m$ at the dominant 816 MHz $TM_{120}$ resonance [28] [29]. From these frequency swept measurements the quality factor of the cell can be estimated [28], $Q= f/df$ ~5. The transverse impedance is proportional to the quality factor, $Q$, of the cell, which is fundamentally the electromagnetic energy stored divided by the energy dissipated. It follows that $Q$ is proportional to the cavity volume and inversely proportional to the surface area and surface resistivity, since the losses are due to resistive heating of the walls. Therefore, if the geometry and wall materials of the ARIA cells are exactly the same as those of the Axis-1 cells, one can also expect the transverse impedance to be exactly the same. Of course, this depends on the wall material being thicker than a skin depth at the $TM_{1n0}$ BBU mode frequencies; not a problem for metal walls, and the ARIA cells incorporate a 1" thick Axis-I ferrite core as the single nonmetallic wall of the cell [14], which is much thicker than the skin depth in ferrite at the 816 MHz resonance (< 1mm).

On DARHT Axis-II, using $I$, $N$, and $\langle 1/B \rangle$, measured during the experiments, and $Z_\perp$ measured on a single cell using two-wire excitation, accurately predicted BBU amplification from Eq. (4) [25]. The focusing field in DARHT Axis-II is almost continuous, so one might expect agreement with the theory, which was developed for a continuous magnetic field. However, unlike Axis-II, the focusing field for Axis-I and ARIA has large voids between 8-cell-blocks to accommodate vacuum pumping and diagnostics. Therefore, the direct calculation of $\langle 1/B \rangle$ from the XTR field may be inappropriate. For example, including the near-zero void field in the average would imply exponential growth in the void, whereas in actuality displacement growth is only linear in a drift region. For this reason, I exclude the void field from the calculation of $\langle 1/B \rangle$, by using a cellblock average for $\langle 1/B \rangle$, ignoring the field if less than 100 Gauss[†].

Moreover, the measurement of transverse impedance on the Axis-II cell used a more advanced technique than was used to measure the Axis-I cell. Therefore, in order to scale BBU growth from Axis-I impedance measurements to ARIA parameters, I applied an empirical scaling factor to $Z_\perp$ in Eq. (5). A scaling factor $\zeta = 1.21$ brings the theory into satisfactory agreement with the data (Fig. 14). The same procedure was then used to predict BBU amplification on ARIA for the two tunes considered herein, based on the strategy of using a cavity design identical to Axis-I. Block averaging should account for the different dimensions of ARIA, and the empirical impedance scaling should also apply to the replicant cavity. Although this analysis is sufficient for *ad hoc* empirical scaling of Axis-I BBU amplification to ARIA, it is clear that a better *ab initio* approach to BBU theory is needed to accommodate quasi-periodic focusing systems like DARHT and ARIA.

---

[†] The use of a 100 G threshold is somewhat arbitrary. However, doubling or halving this value only changed the average by <6% for the low-field tune, < 12% for the high-field tune, and < 20% for Axis-I.

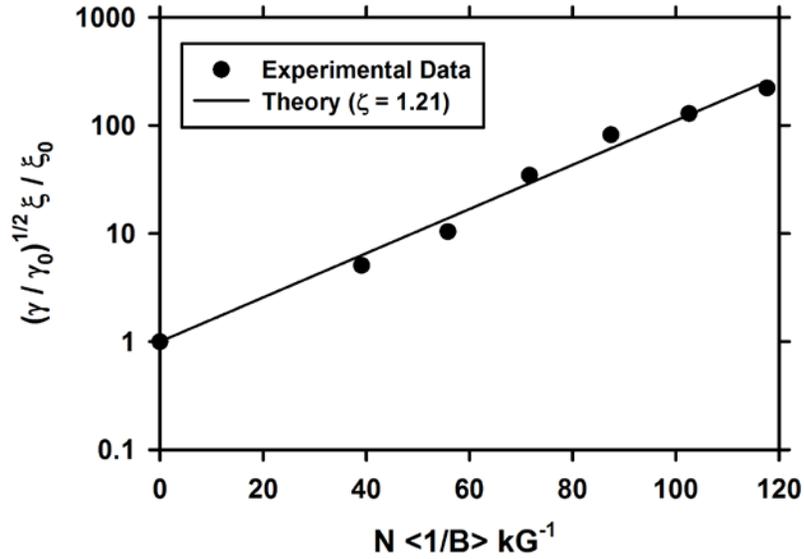

Figure 14: Experimental BBU data [30] fit with the theoretical amplification, using block-averaged $\langle 1/B \rangle$, and empirically scaled transverse impedance (scaling factor $\zeta = 1.21$ ). (*Adapted from ref.* [30].)

The BBU characteristics of these two tunes are compared in Table I, along with the same calculation for the first 36 cells of Axis-I. It is seen that the low-field ARIA tune is somewhat more unstable than the present nominal Axis-I tune, but that the BBU amplification can be reduced by at least a factor of five through application of higher guide fields.

Table I. BBU properties of tunes

|  | $\langle 1/B_z \rangle$ | $\Gamma_{max}$ | $\xi/\xi_0$ |
|---|---|---|---|
|  | kG$^{-1}$ |  |  |
| Low-Field Tune (Fig. 9) | 2.51 | 4.16 | 33.8 |
| High-Field Tune (Fig. 11) | 1.49 | 2.47 | 6.3 |
| DARHT Axis-I (36 cells) | 2.28 | 3.78 | 23.4 |

**B.** *Corkscrew motion*

Strictly speaking, corkscrew motion [31] (or beam sweep [27]) is not an instability. Rather, it is the result of temporal variation of the beam energy interacting with transverse magnet fields in the LIA. The beam deflection by these fields is roughly inversely proportional to beam energy, so time varying beam energy causes time varying deflections that manifest themselves as corkscrew or sweep at the accelerator exit. High-frequency corkscrew is particularly worrisome, because it can seed the BBU.

The amplitude of the corkscrew is approximately [32]

$$A \approx \sqrt{N} \delta\ell \frac{\delta\gamma}{\gamma} \phi_{total} \qquad (7)$$

where $A^2 = \langle x^2 \rangle_t + \langle y^2 \rangle_t$ over a time $t$. Also, $N$ is the number of magnets, $\delta\ell$ is the rms misalignment, and $\phi_{total}$ is the total phase advance ($\int k\, dz$). The cell misalignment includes both offset and tilt, with the tilt contribution approximately the solenoid length times the rms tilt angle (in quadrature with the rms offset). There are two regimes of corkscrew during the beam pulse: on the leading edge and during the flattop.

For the flattop region of the beam pulse, Table II lists the parameters in Eq. (7) for Axis-II, where we successfully suppress corkscrew using our dipole correctors in a few cells [27]. The measured amplitudes of the un-corrected corkscrew are in reasonable agreement with Eq. (7) (see Table II), so this equation can be used as a predictor of the level of corkscrew that can be suppressed. Also listed in this table are the requirements for ARIA, either based on achieved values on DARHT Axis-II [33] [34], or the requirements for DARHT Axis-I [35]. Since the anticipated ARIA corkscrew amplitude is less than Axis-II, we expect to be able to control it in the same way, so long as the cell alignment and voltage variation requirements are met.

Table II: Corkscrew parameters

| Parameter | symbol | unit | Axis-II [27] | ARIA* | ARIA* |
|---|---|---|---|---|---|
| Tune | | | Nominal | Low-Field | High-Field |
| Number of cells | $N$ | | 68 | 36 | 36 |
| RMS offset | $\delta r$ | mm | 0.1 | 0.1 - 0.15 | 0.1 - 0.15 |
| RMS tilt | $\delta\theta$ | mr | 0.3 | 0.3 - 0.65 | 0.3 - 0.65 |
| Solenoid length | $L$ | m | 0.381 | 0.532 | 0.532 |
| RMS Misalignment | $\delta\ell$ | mm | 0.15 | 0.19 - 0.38 | 0.19 - 0.38 |
| Final Energy | $KE$ | MeV | 16.6 | 12 | 12 |
| Energy Variation | $\delta KE / KE$ | % | 2.5 | 2.5 - 1 | 2.5 - 1 |
| Phase Advance | $\phi_{total}$ | radian | $14.52\pi$ | $7.61\pi$ | $15.81\pi$ |
| Amplitude (Eq. (7)) | $A$ | mm | 1.41 | 0.64 - 0.51 | 1.40 - 1.12 |
| Amplitude (measured) | $A_{measured}$ | mm | 1.91 | -- | -- |

\* First numbers in ARIA columns are achieved DARHT Axis-II values.
Second numbers are DARHT Axis-I requirements

Corkscrew during the leading edge of the pulse is more dangerous, because $\delta\gamma/\gamma$ can be large, and if the risetime is short the frequency can be high enough to trigger the BBU. Moreover, control of leading edge corkscrew through the use of DC corrector dipoles and the "Tuning V" algorithm has not been demonstrated. The classic example of violent leading edge corkscrew exciting the BBU was on the ATA accelerator [36], where misalignment was so bad that magnetic transport was soon abandoned in favor of IFR ion-channel guiding.

On the leading edge of the pulse $\delta\gamma/\gamma$ has two contributions, one from beam loading of the cell and the other from the injector. The diode injector produces a current-pulse leading edge with energy rising from zero to the full diode energy. In Axis-I the diode current is essentially zero until the voltage reaches ~ 1.5 MV. Moreover, the earliest part of the leading edge, with energy less than ~3-MeV, begins to be scraped off on the beam pipe before reaching the first cell [37]. Thus, for a 3.7-MV Axis-I diode AK voltage, the beam head injected into the LIA may have $\delta KE < 2.0$ MeV with $\delta\gamma_{rms} < 0.8$ for the first 36 cells. Also on Axis-I, the beam loading is 35-keV per cell, with rms $\delta KE_{rms} = 0.004$ MeV, which is so much less than the injected $\delta KE$ that it can be neglected. The shorter cells of Axis-I make the probable range of misalignments somewhat less than ARIA, 0.14mm to 0.26 mm compared to the ARIA values given in Table II. Using $\delta\ell \approx 0.2$mm, the resulting leading edge amplitude (Eq. (7)) for Axis-I is shown in Fig. 15.

To compare with ARIA, one needs to know the leading edge energy increment ($\delta\gamma$) that is not scraped off before injection. TRAK simulations of the diode suggested that beam energies greater than ~ 1.8 MeV will not be scraped before reaching the first cell block. Assuming that the ARIA hot cathode does not exhibit the delayed turn on apparent on Axis-I, the $\delta KE$ in the beam head could be as great as Axis-I, even though the final diode energy is less. Thus, assuming that ARIA can be aligned as well as DARHT Axis-II, so that $\delta\ell \approx 0.2$mm, the amplitude estimated from Eq. (7) is plotted in Fig. 15. It is seen that the initial ARIA tune is expected to have leading edge corkscrew much greater than that in Axis-I, so it might be a threat for exciting more BBU than in Axis-I. If very high magnetic fields are required to suppress the BBU, as in the high-field ARIA tune, some attention should be paid to the leading edge corkscrew, perhaps by reducing $\delta\gamma$ by scraping off more of the beam head with apertures, as was done on ATA [36].

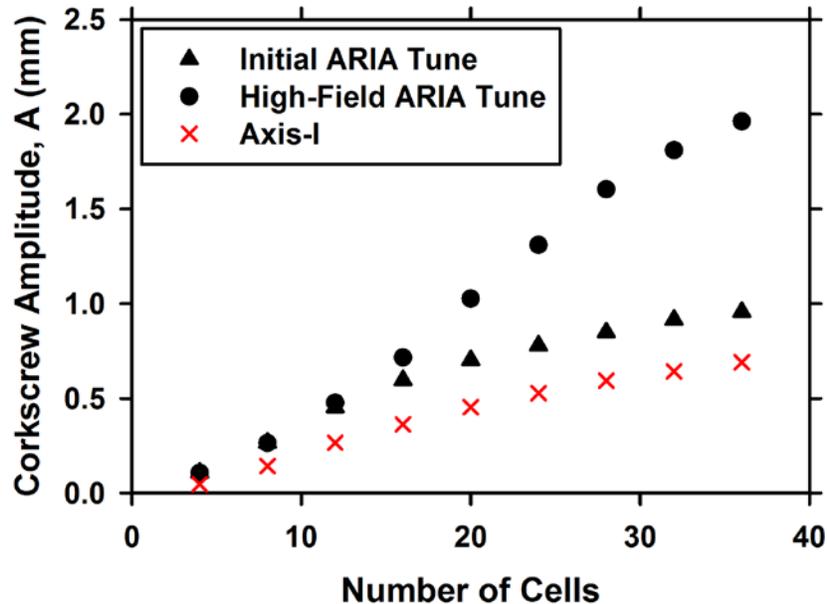

Figure 15: Leading edge corkscrew estimated from Eq. (7) for initial ARIA and high field ARIA tunes, compared with estimates for Axis-I.

## C. *BBU vs Corkscrew*

Once an accelerator is in operation, the most straightforward strategy for reducing the BBU is to increase the magnetic guide field. However, it is sometimes pointed out that doing so increases the corkscrew, which in turn may increase the initial perturbation amplified by the BBU. On the other hand, it is clear from Eq. (4), Eq. (5), and Eq. (7) that as the guide field is increased, corkscrew only grows linearly while BBU is suppressed exponentially. This is illustrated in Fig. 15, where it is seen that BBU is reduced by more than a factor of 5 with only a 50% increase in corkscrew. Thus, increasing the magnetic field to suppress BBU is the most effective strategy, and it expected to be employed on ARIA during commissioning.

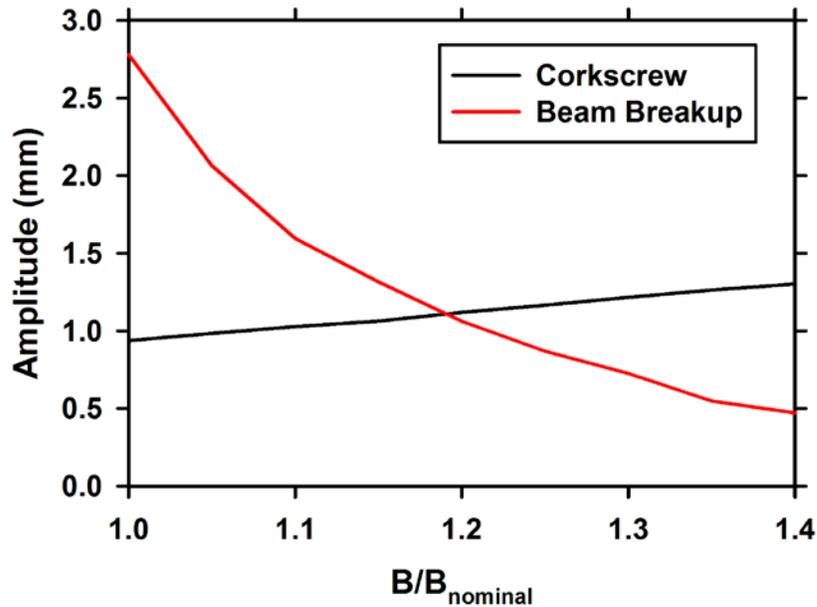

Figure 16: Amplitude of BBU and corkscrew as function of solenoidal focusing field strength for typical ARIA parameters.

## D. *Image Displacement instability*

The image displacement instability (IDI) is also the result of a slightly offset beam interacting with a cavity [38] [39] [40]. Whereas the BBU is the result of specific cavity resonances interacting with the beam, the IDI has no frequency dependence, because it is the result of the difference of magnetic and electric field boundary conditions, so it can perturb the beam even at the lowest frequencies. Moreover, unlike the BBU, the IDI has a definite stability threshold. That is, the beam is unstable in a guide field less than $B_{min}(\gamma, I_b)$, which a function of beam energy, current, and accelerator geometry.

In a beam pipe a slightly offset beam is attracted to the wall by the image of its space charge, and repelled from the wall by the image of its current. These forces balance to within $1/\gamma^2$, with the net force being attractive toward the wall. This is normally counterbalanced by the focusing field. However, in the vicinity of a gap in the wall, the induced charge on the wall collects at the gap edges, and the electric field of the beam decays with radius much more rapidly in the cavity than in the pipe. Thus, if the gap is short compared to the tube radius, the position of the image line charge is almost unchanged. On the other hand, the azimuthal magnetic field of the beam decays with radius exactly as in a pipe with radius equal to the outer wall of the cavity, and the effect is as if the current image was located at a greater distance, reducing the repulsive force from the wall. If the focusing field is too weak, the beam will be displaced toward the wall, and this effect will cumulate as the beam transits each successive gap.

Two recent approaches to IDI theory treat the problem in the limit of a narrow gap for a deep cavity, and neglect the magnetic repulsive force in the vicinity of the gap. Caporaso reduces the equation of motion to the well-known Mathieu equation, which has zones of stability depending on beam parameters [41]. In canonical form, the Mathieu equation is [42]

$$\frac{d^2\psi}{d\zeta^2}+(a-2q\cos 2\zeta)\psi = 0 \tag{8}$$

From the IDI theory the parameters are

$$a = \left(\frac{k_\beta L}{\pi}\right)^2 - \left(\frac{2I}{I_A\beta}\right)\left(\frac{wL}{\pi^2 b^2}\right)$$

$$q = 2\left(\frac{2I}{I_A\beta}\right)\left(\frac{wL}{\pi^2 b^2}\right)\frac{\sin\vartheta}{\vartheta} \tag{9}$$

$$\vartheta = \frac{\pi w}{L}$$

where $w$ is the gap width, $b$ is the tube radius, and $L$ is the intergap spacing. The boundaries of stable solutions of Eq. (8) are well known functions $a_n(q)$. In particular, for parameters relevant to radiography LIAs, $q \ll 1$ and stability obtains for

$$0 \leq a \leq 1 - q - q^2/8 \tag{10}$$

From these considerations, the minimum and maximum focusing field at the ARIA entrance and exit were calculated. In Table III, these are compared with cell-block averages for the low-field and high field tunes, using the magnetic field at the gap locations. Also shown in the table are the minimum field estimates of Briggs, who used a simpler theory, which approximated the disturbance as a wake field effect [40]. In that approach, stability obtains for

$$B_{kG}^2 b_{cm}^2 > 2.7\gamma\left(\frac{w}{L}\right)I_{kA}\eta \tag{11}$$

Using $\eta = 1.5$ for the cavity shape parameter as suggested by Briggs, this estimate of the required focusing field for stability is much more conservative than Caporaso's, and it was used as guidance for tuning the DARHT Axis-II injector.

As seen in Table III., the block average fields from XTR for the two ARIA tunes have fields comfortably greater than the Briggs threshold, and also less than the maximum allowable from Caporaso's analysis. Thus, the IDI is not expected to present a problem for ARIA.

Table III. Stability thresholds for the image displacement instability in ARIA compared with XTR fields averaged over blocks of 4 cells.

|  | Energy | Low-field | High-Field | Briggs | Caporaso | Caporaso |
|---|---|---|---|---|---|---|
|  | $KE$ | $\langle B \rangle$ | $\langle B \rangle$ | $B_{min}$ | $B_{min}$ | $B_{max}$ |
|  | MeV | G | G | G | G | G |
| Cell Block 1 | 3.5 | 266 | 261 | 225 | 107 | 1300 |
| Cell Block 9 | 11.5 | 566 | 1585 | 390 | 185 | 3950 |

**E. *Diocotron instability***

The ARIA diode design used for this investigation has included measures to prevent the injection of a hollow beam, because hollow beams in axial magnetic fields can be diocotron unstable under some conditions [10] [11] [43]. The theory of this instability is well founded and has been validated by numerous experiments with both neutral and non-neutral plasmas and relativistic electron beams. Under some conditions, it is evident on the DARHT Axis-I beam when it is tightly focused by the anode magnet. It would be a troublesome source of beam emittance if present on the ARIA beam under normal operating conditions.

The diocotron is an interchange type of instability caused by sheared rotational velocity in a beam with a radial density profile having an off-axis maximum, as in a hollow beam ("inverted" profile [11])[‡]. In a uniform axial magnetic field, the rotational shear is due to the $\mathbf{E} \times \mathbf{B}$ drift produced by beam space charge, which alters the rigid rotation already present from conservation of canonical angular momentum.

The instability is characterized by strength parameter

$$s = q = \omega_p^2 / \omega_c^2 \tag{12}$$

where $\omega_p^2 = e^2 n_e / \gamma m_e \varepsilon_0$ and $\omega_c = eB / \gamma m_e$. Thus, $s = \gamma n_e m_e / \varepsilon_0 B^2$. Numerical and experimental investigations have shown that high current, hollow beams can be unstable for $s <$ 0.1, depending on the gradient of the current profile, with sharper gradients being the most (13) unstable. Moreover, theory shows that the growth rate of the instability is proportional to $\omega_D / \gamma^2$, where the diocotron frequency is $\omega_D \equiv \omega_p^2 / 2\omega_c$. Thus, lower energy beams are more susceptible to this instability.

For the ARIA beam exiting the diode described in the earlier sections the peak of the anode magnet field is B ~ 420 G, and the envelope radius is ~ 4.2 cm giving s ~ 3, so it should be stable. However, reduction of the current, or increasing the anode magnet strength significantly should be approached with caution, especially if the diode produces a hollowish beam.

---

[‡] Since the diocotron is the result of sheared flow in a medium with a density gradient, it is analogous to the Kelvin-Helmholtz (KH) instability in fluids. The local value of $\omega_D$ is an indicator of the local shear driving the instability.

### F. *Parametric Envelope instability*

The electron beam in the ARIA LIA is transported and focused with solenoidal magnetic fields, tuned to match the beam parameters. If not matched, the beam envelope will exhibit breathing mode oscillations. The magnetic focusing field is generated by periodically spaced solenoids to accommodate accelerating gaps, vacuum pumping, and diagnostics. Under some circumstances, beam transport in a spatially modulated magnetic field can cause a parametric instability of beam envelope oscillations, which in turn could cause halo and emittance growth if amplified enough [44].

The simplest case of solenoidal transport is a beam coasting through a constant axial field. For a given beam energy, current, and emittance, an constant envelope radius can be found by setting the right hand side of Eq. (1) to zero; a so-called matched beam with envelope radius:

$$r_m^2 = \frac{1}{2k_\beta^2}\left[K + \sqrt{K^2 + 4k_\beta^2 \varepsilon^2}\right] \quad (14)$$

However, small perturbations in the initial conditions at injection into the transport field can cause m=0 breathing mode oscillations in the envelope with a characteristic wavelength. By solving the envelope equation for small perturbations about the matched radius, the wavenumber of these oscillations is found to be:

$$k^2 = k_\beta^2 + \frac{K}{r_m^2} + 3\frac{\varepsilon^2}{r_m^4} \quad (15)$$

In a uniform field magnetic field, these are stable, but if the focusing field is periodically modulated, they may be parametrically amplified, especially if the field modulation is in resonance with the natural wavelength. If the field is periodically modulated with wavelength $\Lambda$ (e.g. cell length or magnet spacing), the equation for the envelope perturbations can be reduced to a Mathieu equation like Eq. (8) with parameters

$$\begin{aligned} a &= \left(\frac{kL}{\pi}\right)^2 \\ q &= \left(\frac{k_\beta L}{\pi}\right)^2 \frac{\delta B}{B} \end{aligned} \quad (16)$$

which has well-known parametric regions of instability for which the perturbations will grow, especially for low energy beams in high magnetic fields. For example, Fig. 17 shows the growth of envelope oscillations on a 2-kA, 1.5-MeV beam coasting in a 450 Gauss (average) field having 30% modulation at the 0.62-m ARIA cell length, as calculated by solving the envelope equation.

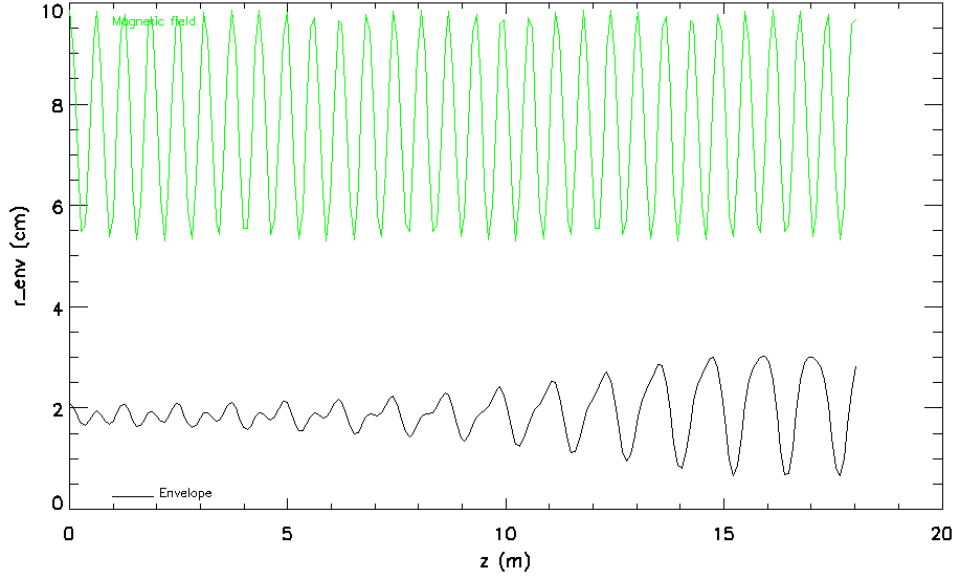

Figure 17: Growth of beam envelope oscillations on a 1.5-MeV, 2-kA beam coasting in a 450 Gauss (average) field having 30% modulation at the 0.62-m ARIA cell length (black curve). The green curve shows the magnetic field in arbitrary units.

A condition that guarantees stability is given by Eq. (10), which is comfortably satisfied by the range of ARIA parameters in the tunes shown in Fig. 10 and Fig. 11. Although the field modulation wavelength $L$ is somewhat longer in ARIA than in DARHT, the solutions are still in the stable region. As before, stability is assured if $0 \leq a < 1 - q - q^2/8$. From Eq. (16) it is clear that $a$ is always positive. The upper bound of this constraint on $a$ sets a maximum allowed guide field, which is tabulated in Table IV for the field modulation depth $\delta B / B \approx 0.3$ apparent in Fig. 9 and Fig. 11. Clearly, any conceivable accelerator tune for ARIA will have fields much less than $B_{max}$, and the parametric envelope instability will not be a problem.

Table IV. Stability threshold for the parametric envelope instability in ARIA.

|  | Energy | Low-field | High-Field | Theory |
|---|---|---|---|---|
|  | $KE$ | $\langle B \rangle$ | $\langle B \rangle$ | $B_{max}$ |
|  | MeV | G | G | G |
| Cell Block 1 | 3.5 | 266 | 261 | 908 |
| Cell Block 9 | 11.5 | 566 | 1585 | 3270 |

Moreover, detailed envelope code simulations show that this instability does not occur in the DARHT or ARIA LIAs for several other reasons, including

- Periodicity of solenoidal field modulation is broken by inter-cellblock gaps.
- Field is changing with distance.
- Field modulation amplitude is spatially varying.

- Beam energy increasing with distance.
- Space-charge and emittance generally force the Mathieu solutions into parametrically stable regions for the parameters of radiographic LIAs.

### G. *Resistive wall instability*

The resistive wall instability is a problem for long-pulse LIAs, but should not be an issue for ARIA. The instability is caused by the beam magnetic field diffusion into the beam-tube wall, whereas the induced charge remains on the surface [45]. Thus, just as for the IDI, the beam is more strongly attracted to the wall. This attraction grows in time as the characteristic magnetic penetration time, so a long pulse beam exhibits a growing head-to-tail displacement. The instability is characterized by a length over which it shows significant growth. In a solenoidal magnetic guide field this length is

$$\Lambda = 5.6 \frac{B_{kG} b_{cm}^3}{I_{kA} \sqrt{\rho_{\mu\Omega-cm} \tau_{\mu s}}} \text{ meters} \tag{17}$$

With growth of an initial perturbation proceeding as $\xi/\xi_0 \approx (\Lambda/z)^{1/3} \exp\left[1.5(z/\Lambda)^{2/3}\right]$. For ARIA parameters in a stainless steel beam pipe the characteristic length is $\Lambda = 587 B_{kG}$ m. So, for ARIA fields greater than the minima required to defeat the IDI (> 0.23 kG), the characteristic length for growth is longer than the accelerator, and this instability should not pose a problem.

### H. *Ion hose instability*

Another instability that can be dangerous for a long pulse accelerator is the ion-hose instability [46]. This is caused by beam-electron ionization of residual background gas. The space-charge of the high-energy beam ejects low-energy electrons from the ionized channel, leaving a positive channel that attracts the beam electrons back if they wander away. This causes the beam to oscillate about the channel position. Likewise, the electron beam attracts the ions, causing them to oscillate about the beam position. Because of the vast differences in particle mass the electron and ion oscillations are out of phase, and the oscillation amplitudes grow.

This instability was of some concern for the long-pulse DARHT Axis-II LIA, and a substantial effort was devoted to understanding it through theory and experiments. The theory of the ion-hose instability in a strong axial guide field such in DARHT-II has been developed in analogy to BBU by treating the ion forces as a continuous transverse impedance [47], and more recently through the use of a spread-mass model [48]. The predictions of these analytic models are in agreement with PIC code simulations [48], including the saturation in time to a maximum growth exponent $\Gamma_m$ in analogy to the BBU. Thus, just as with the BBU, the maximum amplitude should be $\xi(z)/\xi_0 = [\gamma_0/\gamma(z)]^{1/2} \exp\Gamma_m$. From the theory and PIC simulations the maximum growth exponent is

$$\Gamma_m = 0.043 I_{kA} \tau_{\mu s} L_m \left\langle p_{\mu Torr} / \left( B_{kG} a_{cm}^2 \right) \right\rangle \quad (18)$$

where the brackets denote averaging over the LIA length *L*. We experimentally confirmed this on DARHT Axis-II over a wide range of beam parameters in different gasses over a wide range of ion mass [25].

Setting $\Gamma_m \leq 0.693$ for ARIA will ensure that the vacuum is low enough to inhibit the growth of this instability ($\xi/\xi_0 < 1.0$). Using XTR to calculate $\left\langle 1/(Ba^2) \right\rangle$ gives the required vacuum listed in Table IV as maximum allowable pressure. These requirements are easily met. Moreover, using high fields to combat BBU has little effect on vacuum requirements for defeating the ion-hose. Finally it is worth noting that since the ion-hose frequencies are generally less than a few tens of MHz, it would manifest itself as a slow sweep on the 50-ns beam pulse, and would be hard to distinguish from IDI, slow corkscrew, or resistive wall effects.

Table IV. Pressure threshold for 50-ns ARIA pulse.

|  | $I$ | $\tau$ | $L$ | $\left\langle 1/(Ba^2) \right\rangle$ | $p_{max}$ |
|---|---|---|---|---|---|
|  | kA | µs | m | kG$^{-1}$-cm$^{-2}$ | µTorr |
| Low-Field Tune (Fig. 9) | 2 | 0.05 | 24 | 4.49 | 1.5 |
| High-Field Tune (Fig. 11) | 2 | 0.05 | 24 | 3.71 | 1.8 |

## V. ACCELERATOR ENGINEERING DISCUSSION

Beam dynamics considerations have consequences for accelerator engineering details and costs. In general, building on the success of DARHT, if the ARIA engineering and ancillary systems are of the same quality, one can expect the radiography to also be of the same quality. However, there are a number of differences in architecture, which may impact the engineering. In this section, I will attempt to quantify some of these.

### A. *Transport*

Basic transport with beam parameters like the nominal 1.75kA (2" cathode) tune on Axis-I should be no problem. However, much higher magnetic fields would create engineering issues.

It will probably not be possible to use fields as high as those used for the high-field tune discussed here, which was based on using 500A Axis-I power supplies. For example, preliminary calculations indicate that at 450A the temperature rise in the ARIA solenoids would be an uncomfortable 22C [49]. Moreover, simply scaling the Axis-I 500A maximum by input power to the solenoid would give a 400A maximum current for the longer ARIA solenoids.

### B. *BBU*

Limiting the maximum solenoid current to 400A would require a more aggressive ramp up to maximum field than the 500A high-field tune in order to achieve the same low $\langle 1/B \rangle$ as discussed here. To achieve this, the voids in focusing field may need to be bridged with inter-cellblock solenoids or large Helmholtz pairs. Although using Helmholtz pairs may avoid some of the accessibility issues with inter-cellblock solenoids, this would still be an added expense in power supplies and thermal management. Suppression of BBU in ARIA can be better understood through a more rigorous approach to scaling from Axis-I experimental results than the *ad hoc* approach taken here. Since the time-integrated dilation of spot size depends on BBU amplitude as a fraction of equilibrium beam size, this should be the scaling criterion, since both are affected by the strength of the guide field.

**C.** *Corkscrew*

Prevention of large corkscrew during the beam flattop is strictly dependent on the quality of the pulsed power and the care taken to align the magnetic axes of the cells. The amplitude of fully developed corkscrew is proportional to the product of the rms misalignment and the rms energy variation, so one might think of trading off one for the other, especially since we have been so successful at using dipole tuning as mitigation. However, it would be wise to hammer down high frequency noise or ringing on the pulsed power, because high frequency corkscrew can seed the BBU, which is a truly dangerous instability that can only be controlled with strong magnetic fields and the attendant thermal management problems. Flat top energy variation of 1% to 2% (including droop) and alignment to within 0.4mm rms (including both offset and droop) would limit the corkscrew amplitude to less than 1 mm.

**D.** *Image displacement instability*

To prevent the image displacement instability, the average magnetic induction in the first cell block should be greater than 225 G. It should be greater than 390 G in the last cell block. These are based on the conservative Briggs estimates. It is unlikely that the fields will exceed the kGauss maximum values delineated by Caporaso's theory.

**E.** *Diocotron Instability*

The injector for ARIA will be entirely different from the DARHT injectors, and new designs should be approached with caution. The beam out of the diode can be diocotron unstable if it has a hollowish profile and $0.68 \gamma I_{kA} / r_{cm}^2 B_{kG}^2 < 0.1$. The growth rate is $\sim 6 I_{kA} / \gamma^2 r_{cm}^2 B_{kG}$ ns$^{-1}$, so both the threshold and growth rate argue for the highest practical diode voltage. For example, the growth rate on a 1.5 MeV beam is about three times greater than the growth rate on a 3.0 MeV beam. Moreover, care must be taken with diode design in order to prevent generation of hollow profile beams, especially if the beam produced is low energy.

**F.** *Parametric Envelope Instability*

For ARIA beam parameters, the parametric envelope instability is not expected to be a problem, so it has no engineering impact.

**G.** *Resistive Wall instability*

So long as the magnetic guide field in ARIA is greater than the lower limits to be IDI stable, the characteristic growth length of the resistive wall instability will be much longer than the accelerator, so there are no engineering challenges.

**H.** *Ion Hose instability*

The vacuum system must maintain a background pressure less than 1.5 microtorr in order to suppress the ion-hose instability. The present Axis-I system maintains a background much less than 1 micro-torr, so even with the ~50% increase in length (and wall area), using the present Axis-I pumps should prevent this instability. However, detailed vacuum calculations should be performed, and the results used as input to PIC simulations of the ion-hose instability, as was done for Axis-II.

Some of these impacts are summarized in Table VI. The engineering consequences of producing and accelerating a radiographic quality beam with ARIA will obviously be an ongoing discussion as the ARIA design evolves.

Table VI. Beam Physics requirements and Engineering consequences.

| Physics Issue | Accelerator Physics Requirement | Accelerator Engineering Impact |
|---|---|---|
| **Injector:** | | |
| Cathode | Multi-pulse with no AK gap closure | Hot Cathode, Thermal management |
| Diode | Pierce Focusing | Cathode Shroud size |
| Magnetic field | Flux null insensitive to cathode position | External Bucking coil |
| | | |
| **Accelerator:** | | |
| EquilibriumTransport | High-Field tunes | May need Intercellblock solenoids or Helmholtz magnets |
| BBU | ~ kG fields | Magnet Power supplies (400 A for > 20 cells) Thermal Management |
| | Centered beam | Anode steering required |
| Corkscrew | Rms product: | Voltage Pulse Flattop +/- 1% (Axis-I) |
| | $\delta \ell \delta \gamma / \gamma < 4.8$ micron | Cell Alignment: Axis-II or better |
| | Tuning-V implementation | Corrector dipoles in all cells |
| IDI | B > 225 G (CB 1)  B > 390 G (CB 9) | |
| Diocotron | $0.68 \gamma I_{kA} / r_{cm}^2 B_{kG}^2 > 1$  At anode magnet | Diode Voltage  Anode Focusing |
| Ion Hose | p < 1.5 micro-torr | Vacuum system |

## VI. CONCLUSIONS

In general, if the engineering standards used on the DARHT accelerators are adhered to, there should be minimal issues with beam dynamics on ARIA. Of course, commissioning such a machine will involve developing and testing magnetic tunes, including the use of corrector dipoles in some of the cells. However, I do not expect there to be disruptive instabilities or excessive emittance growth, based on the simulations and calculations performed for thie article.

On the other hand, beam breakup in ARIA, and the consequences for accelerator design, deserves a more rigorous treatment than taken here. The scaling herein was based on analytic theories in which the magnetic guide field is either constant, or at least continuous – conditions clearly at odds with the field in Axis-I or ARIA.

Many of the phenomena considered here would result in a slow sweep over the 50-ns beam pulse. It would be hard to distinguish between slow corkscrew resulting from droop in cell or diode voltages, and IDI, resistive-wall instability, or ion-hose instability. However, because the corkscrew increases with magnetic field, whereas the others decrease, increasing the magnetic field is the first thing to try in an effort to reduce a slow sweep. If it turns out to be due to corkscrew (increasing with *B*), then the tuning-V algorithm can be used; it has been shown to be an effective method of corkscrew suppression [27] [44].

Acknowledgements

I am indebted to my colleagues in the DARHT Pulsed-Power and Accelerator Physics group for years of stimulating discourse. Especially noteworthy are discussions I have had with Mark Crawford, Joshua Coleman, Trent McCuistian, David Moir, Kurt Nielsen, Chris Rose, and Martin Schulze.

This research was supported by the National Nuclear Security Administration of the U. S. Department of Energy under contract DE-AC52-06NA25396.

References

[1]  C. Ekdahl, "Modern electron acccelerators for radiography," *IEEE Trans. Plasma Sci.,* vol. 30, no. 1, pp. 254-261, 2002.

[2]  K. Peach and C. Ekdahl, "Particle radiography," *Rev. Acc. Sci. Tech.,* 2013.

[3]  M. Crawford, "ARIA Advanced radiography induction accelerator," Los Alamos National Laboratory report LA-UR-14-20805, 2014.

[4]  S. Humphries, "TRAK: Charged particle tracking in electric and magnetic fields," in *Computational Accelerator Physics*, R. Ryne, Ed., New York, American Institute of Physics, 1994, pp. 597 - 601.

[5]  S. Humphries, "Technical information: TriComp Series," Field Precision, LLC, 2013. [Online]. Available: www.fieldp.com/technical.html.

[6]  "Emission characteristics of "M Type" dipenser cathodes," Heat Wave Labs, Inc. report TB-

117, 2001.

[7] C. Ekdahl, "Diode magnetic-field influence on radiographic spot size," Los Alamos National Laboratory report LA-UR-12-24491, 2012.

[8] H. R. Jory and A. W. Trivelpiece, "Exact relativistic solution for the one-dimensional diode," *J. Appl. Phys.,* vol. 49, pp. 3924 -3926, 1969.

[9] C. Ekdahl, "Axis-1 diode simulations: Standard 2-inch cathode," Los Alamos National Laboratory Report LA-UR-11-00206, 2011.

[10] R. H. Levy, "Diocotron instability in a cylindrical geometry," *Phys. Fluids,* vol. 8, pp. 1288 - 1295, 1965.

[11] R. C. Davidson and G. M. Felice, "Influence of profile shape on the diocotron instability in a non-neutral plasma column," *Phys. Plasmas,* vol. 5, pp. 3497 - 3511, 1998.

[12] C. Ekdahl and et al., "Emittance growth in linear induction accelerators," in *20th Int. Conf. High Power Part. Beams*, Washington, DC, 2014.

[13] C. Ekdahl, "Initial electron-beam results from the DARHT-II linear induction accelerator," *IEEE Trans. Plasma Sci.,* vol. 33, pp. 892 - 900, 2005.

[14] J. Barraza, Los Alamos National Laboratory, Personal communication, 2014.

[15] P. Allison, "Beam dynamics equations for XTR," Los Alamos National Laboratory report, LA-UR-01-6585, 2001.

[16] M. Reiser, Theory and design of charged particle beams, New York. NY: Wiley, 1994, p. p. 467 et seq..

[17] R. L. Gluckstern, "Analytic model for halo formation in high current ion linacs," *Phys. Rev. Lett.,* vol. 73, pp. 1247 - 1250, 1994.

[18] T. P. Wangler and et al., "Particle-core model for transverse dynamics of beam halo," *Phys. Rev. Special Topics - Acc. Beams,* vol. 1, p. 084201, 1998.

[19] T. P. Hughes and et al., "Three-dimensional calculations for a 4-kA, 3.4 MV, 2 microsecond injector with asymmetric power feed," *Phys. Rev. Special Topics - Accel. Beams,* vol. 2, p. 110401, 1999.

[20] C. Thoma and T. P. Hughes, "A beam-slice algorithm for transport of the DARHT-2 accelerator," in *Part. Acc. Conf.*, 2007.

[21] M. Reiser, Theory and design of charged particle beams, New York, NY: Wiley, 1994, pp. 66, et seq..

[22] C. Ekdahl, "Magnetic crosstalk in ferromagnetic materials," Los Alamos National Laboratory Report, LA-UR-14-26009, 2014.

[23] W. K. H. Panofsky and M. Bander, "Asymptotic theory of beam-breakup in linear accelerators," *Rev. Sci. Instrum.,* vol. 39, pp. 206-212, 1968.

[24] V. K. Neil, L. S. Hall and R. K. Cooper, "Further theoretical studies of the beam breakup instability," *Part. Acc.,* vol. 9, pp. 213-222, 1979.

[25] C. Ekdahl and et al., "Long-pulse beam stability experiments on the DARHT-II linear induction accelerator," *IEEE Trans. Plasma Sci.,* vol. 34, pp. 460-466, 2006.

[26] C. Ekdahl and et al., "Beam dynamics in a long-pulse linear induction accelerator," *J. Korean Phys. Soc.,* vol. 59, pp. 3448 - 3452, 2011.

[27] C. Ekdahl and et al., "Suppressing beam motion in a long-pulse linear induction accelerator," *Phys. Rev. ST Accel. Beams,* vol. 14, p. 120401, 2011.

[28] M. Burns, "Cell design for the DARHT linear induction accelerators," in *Particle Accelerator Conference*, 1991.

[29] C. Shang, "BBU design of linear induction accelerator cells for radiographic application," in *Particle Accelerator Conference*, 1997.

[30] J. Coleman and et al., "Increasing the intensity of an induction accelerator and reduction of the beam breakup instability," *Phys. Rev. ST Accel. Beams,* vol. 17, pp. 030101, 1 -11, 2014.

[31] Y.-J. Chen, "Corkscrew modes in linear induction accelerators," *Nucl. Instrum. Methods Phys. Res.,* vol. A292, pp. 455 - 464, 1990.

[32] Y.-J. Chen, "Transverse beam instabilityin a compact dielectric wall induction accelerator," in *Part. Acc. Conf.*, Knoxville, TN, 2005.

[33] H. V. Smith and et al., "X and Y offsets of the 18MeV DARHT-2 accelerator components inside the hall," Los Alamos National Laboratory report LA-UR-09-02040, 2009.

[34] H. V. Smith and et al., "X and Y tilts of the 18MeV DARHT-2 accelerator components," Los Alamos National Laboratory report LA-UR-09-03768, 2009.

[35] "DARHT Facility Work Breakdown Structure Dictionary, 97-D-102," 1998.

[36] G. Caporaso, "Beam dynamics in the Advanced Test Accelerator (ATA)," in *5th Int. Conf. High-Power Particle Beams*, San Francisco, CA, 1983.

[37] J. Coleman and et al., "Characterizing the photon spectrum in the DARHT Axis-I diode," Los Alamos National Laboratory report, LA-UR-14-24384, 2014.

[38] C. H. Woods, "The image instability in high current accelerators," *Rev. Sci. Instrum.,* vol. 41, pp. 959 - 962, 1970.

[39] R. Adler and et al., "The image-displacement effect in intense electron beams," *Part. Acc.,* vol. 13, pp. 25 - 44, 1983.

[40] R. Briggs, "Personal communication," 2005.

[41] G. J. Caporaso and Y.-J. Chen, "Electron Induction Linacs," in *Induction Accelerators*, K. Takayama, Ed., Bearlin, DE, Springer-Verlag, 2011, pp. 117 - 163.

[42] G. Blanch, "Mathieu Functions," in *Handbook of Mathematical Functions*, M. Abramowitz and I. A. Stegun, Eds., Washington, DC, U. S. Government Printing Office, 1965, pp. 722 - 750.

[43] R. B. Miller, An introduction to the physics of intense charged particle beams, New York, NY: Plenum Press, 1982, pp. 126, et seq..

[44] C. Ekdahl, "Tuning the DARHT long-pulse linear induction accelerator," *IEEE Trans. Plasma Sci.,* vol. 41, pp. 2774 - 2780, 2013.

[45] G. Caporaso, W. Barletta and V. K. Neil, "Transverse resistive wall instability of a relativistic electron beam," *Part. Accel.,* vol. 11, pp. 71 - 79, 1980.

[46] H. L. Buchanan, "Electron beam propagation inthe ion-focussed regime," *Phys. Fluids,* vol. 30, pp. 221 - 231, 1987.

[47] R. Briggs, "Transverse instabilities from ion oscillations in the DARHT-II accelerator," Lawence Berkeley National Laboratory report, M7848, 2000.

[48] T. Genoni and T. Hughes, "Ion-hose instability in a long-pulse linear induction accelerator," *Phys. Rev. ST-Accel. Beams,* vol. 6, p. 030401, 2003.

[49] M. Schulze, Los Alamos National Laboratory, personal communication, 2014.

[50] A. G. Cocconi, "Method and apparatus for detecting and preventing impending magnetic saturation in magnetic materials". USA Patent US 4439822, 27 March 1984.

[51] S. Humphries, Charged Particle Beams, New York: Wiley, 1990.